\documentclass[journal,twoside]{IEEEtran}
\usepackage{cite}
\usepackage{amsmath}
\newtheorem{definition}{Definition}
\newtheorem{theorem}{Theorem}
\newtheorem{proposition}{Proposition}
\newtheorem{lemma}{Lemma}

\newtheorem{corollary}{Corollary}
\def \iden {{{I}}}
\def \ux {{{x}}}

\newcommand{\tr}{\rm Tr}
\newcommand{\Trace}{\rm Tr}
\def\reff#1{(\ref{#1})}
\newcommand{\be}{\begin{equation}}
\newcommand{\ee}{\end{equation}}
\newcommand{\bea}{\begin{eqnarray}}
\newcommand{\eea}{\end{eqnarray}}

\begin{document}
\title{Quantum Coding Theorems for Arbitrary Sources, Channels and 
Entanglement Resources}
\author{{Garry~Bowen~and~Nilanjana~Datta}% <-this % stops a space
%\thanks{Manuscript received January 20, 2002; revised November 18, 2002.
   %     This work was supported by the IEEE.}% <-this % stops a space
\thanks{This work was supported by the EPSRC (Research Grant GR/S92816).}%
\thanks{G. Bowen is with the Centre for Quantum Computation, Department of Applied Mathematics and Theoretical Physics, University of Cambridge, Cambridge CB3 0WA, UK (e-mail: gab30@damtp.cam.ac.uk).}
\thanks{N. Datta is with the Statistical Laboratory, Department of Pure Mathematics and Mathematical Statistics, University of Cambridge, Cambridge CB3 0WA, UK (e-mail: N.Datta@statslab.cam.ac.uk).}%
}
% note the % following the last \IEEEmembership and also the first \thanks -
% instance, "A\textbf{} \textbf{}B" will typeset as "A B" not "AB". If you want
% "AB" then you have to do: "A\textbf{}\textbf{}B"
\markboth{Submitted to IEEE Transactions on Information Theory}{Bowen \& Datta: Quantum Information Spectrum Stuff}
% *** Note that you probably will NOT want to include the author's name in ***
% *** the headers of peer review papers.                                   ***

% If you want to put a publisher's ID mark on the page
% (can leave text blank if you just want to see how the
% text height on the first page will be reduced by IEEE)
% \pubid{0000--0000/00\$00.00~\copyright~2002 IEEE}

%\specialpapernotice{(Invited Paper)}

\maketitle

\begin{abstract}
The information spectrum approach gives general formulae for optimal rates 
of various information theoretic protocols, under minimal assumptions
on the nature of the sources, channels and entanglement resources involved. 
This paper 
culminates in the derivation of the 
dense coding capacity for a noiseless quantum 
channel, assisted by arbitrary shared entanglement, using this approach. 
We also review 
the currently known coding theorems, and their converses, 
for protocols such as data compression for arbitrary quantum sources and
transmission of classical information through arbitrary quantum channels. In addition, 
we derive the optimal rate of data compression for a mixed source.
\end{abstract}


% Will be ignored for other modes.
\IEEEpeerreviewmaketitle

\section{Introduction}
% \PARstart{A}{demo} file is ....
% \PARstart{A}{}demo file is ....
% \PARstart{T}{his demo} file is ....
%\PARstart{T}{his} demo file is intended to serve as a ``starter file"
%for IEEE journal papers produced under \LaTeX\ using IEEEtran.cls version
%1.6b and later.

\PARstart{Q}{uantum} information theory generalizes the ideas of coding and
communication to include the nature of the physical system in which information
is encoded.  The information spectrum approach of Han \& Verdu
\cite{verdu94,han} gives general formulae for many operational schemes in
information theory.  It replaces the idea of typical
events (generally called typical sequences) in information theory, with high
probability events.  The power of this approach lies in the lack of
assumptions about the source, channel and entanglement resource.

The quantum information spectrum was defined in terms of quantum states by
Hayashi \& Nagaoka \cite{hayashi03}, initially in the context of hypothesis
testing, and was used to determine a general expression for the classical 
capacity of arbitrary quantum channels.  The quantum information spectrum 
extends the idea of high probability events to high probability subspaces 
of states in a Hilbert space.  In the commutative case, the quantum 
information spectrum simply reduces to its classical counterpart.

In this paper we present a review of coding theorems for quantum data
compression and transmission of classical information through a 
quantum channel.  The rate of
compression for a mixed source is explicitly derived.  A number of new
results are also presented, including the dense coding capacity for a noiseless
quantum channel, assisted by arbitrary shared entanglement.

\section{Preliminaries}

Let ${\cal B}({\cal H})$ denote the algebra of linear operators acting on
a finite--dimensional Hilbert space ${\cal H}$ of dimension $d$. The von Neumann entropy of a state $\rho$, i.e. a positive operator of 
unit trace in ${\cal B}({\cal H})$, is defined as $S(\rho) = -\Trace \rho \log \rho$. Throughout this paper, we choose the logarithm to base $e$. We could
equally well choose an arbitrary base for
the logarithm. This would simply scale the unit of information.

A quantum channel is given by a
completely positive trace--preserving (CPTP) map $\Phi: {\cal
B}({\cal K}) \to {\cal B}({\cal H})$, where ${\cal K}$ and ${\cal
H}$ are the input and output Hilbert spaces of the channel.

\subsection{Spectral Projections}
The quantum information spectrum approach requires the extensive use of
spectral operators.  For a self-adjoint operator $A$ written in its spectral
decomposition $A = \sum_i \lambda_i |i\rangle \langle i|$ we define the
positive spectral projection on $A$ as
\begin{equation}
\{ A \geq 0 \} = \sum_{\lambda_i \geq 0} |i\rangle \langle i|
\end{equation}
the projector onto the eigenspace of positive eigenvalues of $A$.
Corresponding definitions apply for the other spectral projections $\{ A < 0
\}$, $\{ A > 0 \}$ and $\{ A \leq 0 \}$.  For two operators $A$ and $B$, we can
then define $\{ A \geq B \}$ as $\{ A - B \geq 0 \}$, and similarly for the
other ordering relations.

\subsection{Two Important Lemmas}

The following key lemmas are used repeatedly in the paper. For their proofs
see \cite{BD1}.

\begin{lemma}
\label{lemma}
For self-adjoint operators $A$, $B$ and any positive operator $0 \leq P \leq I$
the inequality
\begin{equation}
\mathrm{Tr}\big[ P(A-B)\big] \leq \mathrm{Tr}\big[ \big\{ A \geq B \big\}
(A-B)\big]
\label{eqn:first_ineq}
\end{equation}
holds.
\end{lemma}

\begin{lemma}
\label{lemma2}
For self-adjoint operators $A$ and $B$, and any completely positive
trace-preserving (CPTP) map $\mathcal{T}$ the inequality
\begin{equation}
\mathrm{Tr}\big[ \{\mathcal{T}(A) \geq \mathcal{T}(B) \}\mathcal{T}(A-B)\big]
\leq \mathrm{Tr}\big[ \big\{ A \geq B \big\} (A-B)\big]
\label{eqn:second_ineq}
\end{equation}
holds.
\end{lemma}

We also make use of the following proposition
\begin{proposition}
\label{cor0}
Given a state $\rho_n$ and a self-adjoint
operator $\omega_n$, we have
\be
\mathrm{Tr}\big[\{\rho_n \ge e^{n\gamma}\omega_n \} \omega_n \bigr]
\le e^{-n\gamma}.
\ee
for any real $\gamma$.
\end{proposition}
\begin{proof}
We have
\be
\mathrm{Tr}\big[\{\rho_n \ge e^{n\gamma}\omega_n \} (\rho_n - e^{n\gamma}\omega_n \bigr]
\ge 0
\ee
and hence, by rearranging terms
\be
\mathrm{Tr}\big[\{\rho_n \ge e^{n\gamma}\omega_n \} \omega \bigr]
\le e^{-n\gamma} \mathrm{Tr}\big[\{\rho_n \ge e^{n\gamma}\omega_n \} \rho_n \bigr]
\le e^{-n\gamma}.
\ee
where $\mathrm{Tr}\big[\{\rho_n \ge e^{n\gamma}\omega_n \} \rho_n \bigr] \leq 1$.
\end{proof}

\subsection{Quantum Spectral Information Rates}

As a generalization of the relative entropy, the spectral divergence allows information theory to include arbitrary sources and channels.
\begin{definition}
Given the difference operator $\Pi_n(\gamma) = \rho_n - e^{n\gamma}\omega_n$, the quantum spectral sup-(inf-)divergence rates are defined as
\begin{align}
\overline{D}(\rho \| \omega) &= \inf \Big\{ \gamma : \limsup_{n\rightarrow \infty} \mathrm{Tr}\big[ \{ \Pi_n(\gamma) \geq 0 \} \Pi_n(\gamma) \big] = 0 \Big\} \label{supdiv} \\
\underline{D}(\rho \| \omega) &= \sup \Big\{ \gamma : \liminf_{n\rightarrow \infty} \mathrm{Tr}\big[ \{ \Pi_n(\gamma) \geq 0 \} \Pi_n(\gamma) \big] = 1 \Big\} \label{infdiv}
\end{align}
respectively.
\end{definition}
The spectral entropies, conditional spectral entropies, and spectral mutual
information rates may all be expressed as a divegrence rate with appropriate
substitutions for the sequence of operators $\omega = \{ \omega_n
\}_{n=1}^{\infty}$.  These are
\begin{align}
\overline{S}(\rho) &= -\underline{D}(\rho| I) \label{supent} \\
\underline{S}(\rho) &= -\overline{D}(\rho| I) 
\end{align}
and for sequences of bipartite state $\rho^{AB} = \{\rho^{AB}_n\}_{n=1}^\infty$,
\begin{align}
\overline{S}(A|B) &= -\underline{D}(\rho^{AB}| I^{A}\otimes \rho^B) \\
\underline{S}(A|B) &= -\overline{D}(\rho^{AB}| I^{A}\otimes \rho^B) \\
\overline{S}(A:B) &= \overline{D}(\rho^{AB}| \rho^{A}\otimes \rho^B)\\
\underline{S}(A:B) &= \underline{D}(\rho^{AB}| \rho^{A}\otimes \rho^B),
\end{align}
giving all the spectral sup(inf)-information rates.  Various properties and
relationships of these quantities are explored in \cite{BD1}.

\section{Data Compression for Arbitrary Quantum Sources}

A general quantum source consists of a sequence of density
$\rho = \{\rho_n\}_{n=1}^\infty$ acting on a corresponding
sequence of Hilbert spaces ${\cal{H}} = \{{\cal{H}}_n\}_{n=1}^\infty$.

A compression scheme for such a source, $\rho$, consists of two families
of quantum operations ${\cal{C}}_n$ and ${\cal{D}}_n$. Here
${\cal{C}}_n$ denotes the compression
operation which takes states in the original Hilbert space ${\cal{H}}_n$
to states in a Hilbert space ${\widetilde{{\cal{H}}_n}}$ such that
${\hbox{dim }} {\widetilde{{\cal{H}}_n}} \leq {\hbox{dim }} {{{\cal{H}}_n}}$.
Hence, ${\widetilde{{\cal{H}}_n}}$ can be regarded as the compressed
Hilbert space.
The corresponding decompression operation, ${\cal{D}}_n$, takes states
in ${\widetilde{{\cal{H}}_n}}$ to states in the original Hilbert space
${\cal{H}}_n$.

The compression scheme given by the family of combined
compression decompression maps ${\cal{D}}_n \circ {\cal{C}}_n$ is said to be
\textit{reliable}
if the entanglement fidelity $F(\rho_n, {\cal{D}}_n \circ {\cal{C}}_n)$
tends to $1$ as $n \rightarrow \infty$.  Let $P_n$ denote the orthogonal
projection onto ${\widetilde{{\cal{H}}_n}}$.
The \textit{rate} of the compression scheme is determined by
\begin{equation}
R = \limsup_{n\rightarrow \infty} \frac{1}{n} \log M_n,
\end{equation}
where $M_n := {\tr} P_n = \dim {\widetilde{{\cal{H}}_n}}$.

The objective is thus to obtain the optimal rate of reliable compression for a
given source $\rho$.  Defining the optimal rate $\mathcal{R}$ as the infimum of
all reliable rates, leads to the following theorem.

\begin{theorem}
\label{coding}
The quantum spectral sup-entropy rate is optimal.  Hence,
\begin{equation}
\mathcal{R} = \overline{S}(\rho)
\end{equation}
for a given source $\rho$. Equivalently, $(i)$ if $R> \overline{S}(\rho)$
then there exists a reliable compression scheme of rate $R$, and
$(ii)$ there can be no reliable compression scheme of
rate $R$ for $R< \overline{S}(\rho)$.
\end{theorem}

\begin{proof}[Proof of (i) :]
Suppose $R> \overline{S}(\rho)$.
Consider the compression operation, ${\cal{C}}_n$, defined by its action on
any state $\sigma_n \in {\cal{B}}({\cal{H}}_n)$ as follows:
\be
{\cal{C}}_n(\sigma_n):= P_n\sigma_nP_n + \sum_{k} A_k \sigma_n A_k^\dagger,
\ee
where $(a)$ $P_n$, the compression projection, i.e. the orthogonal projection
onto
the compressed Hilbert
space ${\widetilde{{\cal{H}}_n}}$, is given by
\be
P_n:= \{\rho_n \ge e^{-n\gamma} I_n \},
\ee
and $(b)$ $A_k := |\chi_0\rangle \langle k|$, with $|\chi_0\rangle$
being a fixed pure state in ${\widetilde{{\cal{H}}_n}}$ and
$\{|k\rangle\}$ being an orthonormal basis for the orthocomplement of
${\widetilde{{\cal{H}}_n}}$. Equivalently,
\be
{\cal{C}}_n(\sigma_n):= P_n\sigma_nP_n + {\tr}\bigl((I_n - P_n)\sigma_n\bigr)
| \chi_0\rangle \langle\chi_0 |.
\ee
The corresponding decoding operation ${\cal{D}}_n$ is defined to be the
identity
on ${\widetilde{{\cal{H}}_n}}$.

If $\{C_n^j\}$ and $\{D_n^k\}$ denote
finite sets of Kraus operators of the quantum
operations ${\cal{C}}_n$ and ${\cal{D}}_n$ respectively, then
\be
F_n:= F(\rho_n, {\cal{D}}_n \circ {\cal{C}}_n) =
\sum_{jk} |{\tr}\bigl(D_n^k C_n^j \rho_n)|^2.
\label{entfid}
\ee
and hence the entanglement fidelity is given by
\begin{eqnarray}
F(\rho_n, {\cal{D}}_n \circ {\cal{C}}_n) &=&
| {\tr}(P_n \rho_n )|^2 + \sum_k |{\tr}( A_k \rho_n )|^2 \nonumber\\
&\ge & |{\tr}(P_n \rho_n )|^2 \nonumber\\
&\ge & |{\tr}\bigl[P_n (\rho_n - e^{-n\gamma} I_n)\bigr]|^2 \nonumber\\
&= &|{\tr}\bigl[\{\rho_n \ge e^{-n\gamma} I_n \}(\rho_n - e^{-n\gamma}
I_n)\bigr]|^2\nonumber\\
&= &|{\tr}\bigl[\{ \Pi_n(\gamma) \geq 0 \} \Pi_n(\gamma)\bigr]|^2,\nonumber\\
\label{fidlim}
\end{eqnarray}
where $\Pi_n(\gamma) = \rho_n - e^{n\gamma}I_n$.
From the definitions in (\ref{supent}) and (\ref{infdiv}) it follows that
the RHS of \reff{fidlim} tends to $1$ as $n \rightarrow \infty$, for any
$\gamma > \overline{S}(\rho)$.

Utilizing Proposition \ref{cor0}, the dimension of the compression projections
$P_n$ is bounded for each $n$
by
\begin{equation}
\mathrm{Tr} P_n = \mathrm{Tr} \big[ \{ \rho_n \geq e^{-n\gamma} I_n \} \big]
\leq
e^{n\gamma} = e^{n(\overline{S}(\rho) + \delta)}
\end{equation}
for $\delta > 0$.  Since this is true for all $\delta > 0$ we have $\mathcal{R}
\leq
\overline{S}(\rho)$.

[Proof of (ii) (Weak Converse):]  Suppose $R < \overline{S}(\rho)$.
Without loss of generality, assume that ${\cal{C}}_n$ maps states in
${\cal{H}}_n$ to states in an $M_n$-dimensional Hilbert space
 ${\widetilde{{\cal{H}}_n}}$, with $M_n = \lfloor e^{nR}\rfloor$.  Hence, if
$P_n$ is the orthogonal projection onto ${\widetilde{{\cal{H}}_n}}$ then
${\tr} [P_n] = M_n \le e^{nR}$.

Let $\{C_n^j\}$ and $\{D_n^k\}$ denote
finite sets of Kraus operators for the quantum
operations ${\cal{C}}_n$ and ${\cal{D}}_n$ respectively. Obviously,
$P_n C_n^j = C_n^j$. Further, let $Q_n^k$ be the orthogonal
projection onto the subspace to which
${\widetilde{{\cal{H}}_n}}$ is mapped to by $D_n^k$. Then
$D_n^kC_n^j = D_n^kP_n C_n^j
= Q_n^k D_n^kP_n C_n^j = Q_n^k D_n^k C_n^j$. Moreover,
${\tr} [Q_n^k ] \le {\tr} [P_n]$ since ${\cal{D}}_n$ is a CPTP map.

The entanglement fidelity can be
expressed as
\begin{align}
F_n &= \sum_{jk} |{\tr}\bigl(D_n^k C_n^j \rho_n)|^2\nonumber\\
&=  \sum_{jk} |{\tr}\bigl(Q_n^k D_n^k C_n^j \rho_n)|^2 \nonumber\\
&=  \sum_{jk} |{\tr}\bigl[ (D_n^k C_n^j \sqrt{\rho_n})(\sqrt{\rho_n}Q_n^k)
\bigr]|^2\nonumber\\
&\le \sum_{jk} {\tr} \bigl[Q^k_n{\rho_n}Q_n^k\bigr]\cdot {\tr}\bigl[ D_n^k
C_n^j {\rho_n}C_n^{j\dag}D_n^{k\dag}\bigr] \label{long}\\
&\le {\tr}\bigl[P_n \rho_n\bigr] \label{long2}\\
&\le {\tr}\bigl[\{\rho_n \ge e^{-n\gamma}I_n\}(\rho_n - e^{-n\gamma}I_n)\bigr]
\nonumber \\
&\phantom{=}\;+  e^{-n\gamma}{\tr} P_n
\label{long1}
\end{align}
To arrive at \reff{long}, we have made use of the Cauchy Schwarz inequality for
the
Hilbert-Schmidt inner product, specifically $|{\tr}(A^\dagger B)|^2 \le
{\tr}(A^\dagger A)
\cdot {\tr}(B^\dagger B)$.  The inequality in \reff{long2} uses the inequality
${\tr} Q_n^k\le {\tr} P_n$, and the fact that $\mathcal{C}_n$ and
$\mathcal{D}_n$ are trace preserving maps. The final inequality in \reff{long1}
follows from Lemma \ref{lemma}.

Using the fact
that
${\tr} P_n \le e^{nR}$, we have
\be
F_n \leq {\tr}\bigl[\{\rho_n \ge e^{-n\gamma}I_n\}(\rho_n -
e^{-n\gamma}I_n)\bigr]
+ e^{-n(\gamma - R)}.
\label{last}
\ee
Choosing a number $\gamma$ and $\delta > 0$ such that $R = \gamma + \delta <
\overline{S}(\rho)$, the second term on RHS of \reff{last} tends to zero as $n
\rightarrow \infty$. However, since $\gamma < {\overline{S}}(\rho)$
the first term on RHS of \reff{last} does not converge to $1$ as
$n \rightarrow \infty$. Hence, the fidelity does not converge to 1 in the limit
as $n \rightarrow \infty$ and the compression scheme is not reliable.
\end{proof}

The proof of the weak converse above shows that for $R < \overline{S}(\rho)$
the entanglement fidelity cannot approach unity, and hence any compression
scheme will give an error with non-zero probability.  To determine the rate at
which the probability of error converges to 1 for any compression protocol we
can equivalently determine the supremum of the rates for which the asymptotic
limit of the entanglement fidelity goes to zero.  Here we prove the result for
the \textit{strong converse} rate denoted by $\mathcal{R}^*$.
\begin{theorem}
Coding a source $\rho$ at a rate less than the quantum inf-spectral entropy
rate gives an error with probability equal to one.  That is
\begin{equation}
R < \underline{S}(\rho) \implies \lim_{n\rightarrow \infty}F_n = 0.
\end{equation}
or, equivalently $\mathcal{R}^* = \underline{S}(\rho)$.
\end{theorem}
\begin{proof}
From \reff{long1} we can immediately see that for rates $R <
\underline{S}(\rho)$ choosing a $\gamma = R + \delta < \underline{S}(\rho)$ we
obtain
\begin{equation}
\lim_{n\rightarrow \infty} F_n  = 0
\end{equation}
and the compression scheme fails with probability approaching 1 as $n
\rightarrow \infty$.
\end{proof}

\subsection{Relationship to the von Neumann Entropy}

For any quantum information source $\rho$, the quantum spectral sup- and inf-
information rates are related to the von Neumann entropy in the following
manner.

\begin{lemma}
The sup-information and inf-information rates are related to the von Neumann
entropy by
\begin{equation}
\underline{S}(\rho) \leq \liminf_{n\rightarrow \infty} \frac{1}{n}S(\rho_n)
\leq \limsup_{n\rightarrow \infty} \frac{1}{n}S(\rho_n) \leq \overline{S}(\rho)
\end{equation}
for any source $\rho$.
\end{lemma}
\begin{proof} Let $\{\lambda_n^i\}$ denote the set of eigenvalues of
state $\rho_n$.
For the first inequality we have
\begin{align}
\frac{1}{n}S(\rho_n) &= - \frac{1}{n} \mathrm{Tr} \big[ \rho_n \log \rho_n
\big] \nonumber \\
&= - \frac{1}{n} \sum_i \lambda^i_n \log \lambda^i_n \nonumber \\
&\geq - \frac{1}{n} \sum_{\lambda_n^i < e^{-n(\underline{S}(\rho) -
\delta)}} \lambda^i_n \log \lambda^i_n \nonumber \\
&\geq - \frac{1}{n} \mathrm{Tr}\big[ \{ \rho_n < e^{-n(\underline{S}(\rho) -
\delta)} \} \rho_n \big] \log e^{-n(\underline{S}(\rho) - \delta)} \nonumber \\
&= (\underline{S}(\rho) - \delta)\mathrm{Tr}\big[ \{ \rho_n <
e^{-n(\underline{S}(\rho) - \delta)} \} \rho_n \big]
\end{align}
and from the definition of $\underline{S}(\rho)$ we have $\lim_{n \rightarrow
\infty} \mathrm{Tr}\big[ \{ \rho_n \leq e^{-n(\underline{S}(\rho) - \delta)} \}
\rho_n \big] = 1$, and this is true for all $\delta > 0$, implying
\begin{equation}
\underline{S}(\rho) \leq \liminf_{n\rightarrow \infty} \frac{1}{n}S(\rho_n)
\end{equation}
Similarly, we have
\begin{align}
\frac{1}{n}S(\rho_n) &= - \frac{1}{n} \sum_{\lambda_n^i \geq
e^{-n(\overline{S}(\rho) + \delta)}} \lambda^i_n \log \lambda^i_n \nonumber \\
&\phantom{=}\qquad \quad - \frac{1}{n} \sum_{\lambda_n^i <
e^{-n(\overline{S}(\rho) + \delta)}} \lambda^i_n \log \lambda^i_n \nonumber \\
&\leq (\overline{S}(\rho) + \delta)\mathrm{Tr}\big[ \{ \rho_n
\geq e^{-n(\overline{S}(\rho) + \delta)} \} \rho_n \big] \nonumber \\
&- \frac{1}{n} \mathrm{Tr}\big[Q_n \rho_n \log \, \rho_n\bigr],
\label{eq11}
\end{align}
where $Q_n := \{ \rho_n < e^{-n(\overline{S}(\rho) + \delta)} \}$. Let
$W_n := Q_n \rho_n Q_n$ and define the normalized state ${\widehat{W}}_n := W_n/(
 \mathrm{Tr} W_n)$. Hence,
\begin{align}
 \frac{1}{n}S(\rho_n)&\leq (\overline{S}(\rho) + \delta)\mathrm{Tr}\big[ \{ \rho_n
\geq e^{-n(\overline{S}(\rho) + \delta)} \} \rho_n \big] \nonumber \\
&- \frac{1}{n} \mathrm{Tr} W_n \bigl(\log \, W_n + \log  \mathrm{Tr} W_n 
- \log  \mathrm{Tr} W_n \bigr)\nonumber\\
&= (\overline{S}(\rho) + \delta)\mathrm{Tr}\big[ \{ \rho_n
\geq e^{-n(\overline{S}(\rho) + \delta)} \} \rho_n \big] \nonumber \\
&- \frac{1}{n} \mathrm{Tr} W_n S({\widehat{W}}_n) - \frac{1}{n} H(\mathrm{Tr} W_n),
\nonumber\\
&\leq (\overline{S}(\rho) + \delta)\mathrm{Tr}\big[ \{ \rho_n
\geq e^{-n(\overline{S}(\rho) + \delta)} \} \rho_n \big] \nonumber \\
&+ \frac{1}{n} \log d_n\, \mathrm{Tr} W_n - \frac{1}{n} H(\mathrm{Tr} W_n)
\end{align}
In the above, $H(\cdot)$ denotes the Shannon entropy. 
and $d_n={\hbox{dim\,}}\mathcal{H}_n$.
Since $\lim_{n \rightarrow \infty} \mathrm{Tr} W_n = \lim_{n \rightarrow \infty} \mathrm{Tr}\big[ \{ \rho_n <
e^{-n(\overline{S}(\rho) + \delta)} \} \rho_n \big] = 0$, the last term 
vanishes in this limit. The second term also vanishes under the
assumption that for all $n$
\begin{equation}
\frac{1}{n} \log d_n < \beta
\end{equation}
for some $\beta < +\infty$. Moreover, since
$\lim_{n\rightarrow \infty} \mathrm{Tr}\big[ \{ \rho_n \geq
e^{-n(\overline{S}(\rho) + \delta)} \} \rho_n \big] = 1$, we have
\begin{equation}
\limsup_{n\rightarrow \infty} \frac{1}{n}S(\rho_n) \leq \overline{S}(\rho).
\end{equation}
The remaining inequality follows from the definition of $\liminf$ and
$\limsup$.
\end{proof}

\subsection{Mixed Sources}

Given two sources $\sigma= \{\sigma_n\}_{n=1}^\infty$ and
$\omega=\{\omega_n\}_{n=1}^\infty$, we define the mixed source
$\rho=\{\rho_n\}_{n=1}^\infty$ to be the
source for which
\begin{equation}
\rho_n = t \sigma_n + (1-t) \omega_n
\end{equation}
for $t \in (0,1)$.
\begin{theorem}
For the mixed source $\rho$ the optimal rate $\mathcal{R}$ is given by
\begin{equation}
\mathcal{R} = \max \big[ \overline{S}(\sigma), \overline{S}(\omega)\big],
\end{equation}
the maximum of the rates for either source $\sigma$ or $\omega$.
\end{theorem}
\begin{proof}
Let $\mathrm{Tr}\big[\Pi_n(\gamma)\big] = \mathrm{Tr}\big[ \{ \rho_n \geq e^{-n\gamma}I_n \} (\rho_n - e^{-n\gamma}I_n)\big]$, then from the linearity of the trace operation, we have
\begin{align}
\mathrm{Tr}\big[\Pi_n(\gamma)\big] &= t\, \mathrm{Tr}\big[ \{ \rho_n \geq e^{-n\gamma}I_n \} (\omega_n - e^{-n\gamma}I_n)\big] \nonumber \\
&\phantom{=}\; + (1-t)\, \mathrm{Tr}\big[ \{ \rho_n \geq e^{-n\gamma}I_n \} (\sigma_n - e^{-n\gamma}I_n)\big] \nonumber \\
&\leq t \, \mathrm{Tr}\big[ \{ \omega_n \geq e^{-n\gamma}I_n \} (\omega_n - e^{-n\gamma}I_n)\big] \nonumber \\
&\phantom{=}\; + (1-t) \, \mathrm{Tr}\big[ \{ \sigma_n \geq e^{-n\gamma}I_n \} (\sigma_n - e^{-n\gamma}I_n)\big]
\label{mixed1}
\end{align}
where the inequality follows from Lemma \ref{lemma}.  Hence for any $\gamma = \overline{S}(\rho) + \delta$, the limit of the LHS goes to one, and hence both of the traces on the RHS must also approach one in the limit.  This then implies that
% \begin{equation}
% \lim_{n \rightarrow \infty} \mathrm{Tr} \big[ P_n (t\sigma_n + (1-t)\omega_n)
% \big] = 1
% \end{equation}
% implies that both
% \begin{equation}
% \lim_{n \rightarrow \infty} \mathrm{Tr} \big[ P_n \sigma_n \big] = 1
% \label{eqa}
% \end{equation}
% and
% \begin{equation}
% \lim_{n \rightarrow \infty} \mathrm{Tr} \big[ P_n \omega_n \big] = 1.
% \label{eqb}
% \end{equation}

% Theorem \ref{coding} $(i)$ tells us that a sequence of projection operators
% $\{P_n\}_{n=1}^\infty$ for which
% $$ {{\tr}}\bigl[ P_n \rho_n\bigr] \rightarrow 1 \quad {\hbox{as}}\quad n
% \rightarrow \infty,$$
% results in a reliable compression scheme for the source $\rho =
% \{\rho_n\}_{n=1}^\infty$, with $(i)$ $P_n$ being the compression projections;
% $(ii)$ the
% decoding operation being the identity on the compressed Hilbert Space,
% and $(iii)$  ${{\tr}}P_n \leq e^{n\overline{S}(\rho )}$.
% Hence, \reff{eqa} and
% \reff{eqb} imply that $P_n$ defined above must be reliable compression
% projections for the sources $\sigma$ and $\omega$ as well.  Hence,
% $$
% {{\tr}}P_n \geq e^{n\overline{S}(\sigma )} \quad {\hbox{and}} \quad {{\tr}}P_n \geq
% e^{n\overline{S}(\omega )}.
% $$
% from the definition of the rate
% $\mathcal{R}$ and the statement of Theorem \ref{coding}.  It follows that
\begin{equation}
\overline{S}(\rho ) \geq \max \big[ \overline{S}(\sigma), \overline{S}(\omega)
\big]
\end{equation}
as $\delta$ is arbitrary.
% follows from Theorem \ref{coding}.

%%%%%%%%%%% end insertion %%%%%%%%%%%%%555
To prove the reverse inequality we explicitly construct a sequence of
projection operators. For each $\alpha>0$ we utilize the projections
$P_n^0:=\{\sigma_n \geq e^{-n\alpha}I_n \}$ and $Q_n:=\{\omega_n \geq e^{-n\alpha}I_n
\}$. Let $Q_n$ have the spectral projection $Q_n = \sum_{i=1}^K |i\rangle
\langle i|$,
with $K= {\tr} Q_n$. Starting with $P_n^0$, we define a sequence of projection
operators $P_n^i$, $i=1, \ldots, K$, iteratively, as follows.
For each $i$, if $|i\rangle$ lies in the subspace onto which $P_n^{i-1}$
projects,
then we set $P_n^i = P_n^{i-1}$.  Otherwise, we take the component of
$|i\rangle$ orthogonal
to this subspace, say $|i^{\perp}\rangle$, and let $P_n^i = P_n^{i-1} \oplus
|i^{\perp}\rangle \langle i^{\perp}|$.

From Lemma 1 it then follows that
\begin{align}
\mathrm{Tr}\big[\Pi_n(\gamma)\big] &\geq \mathrm{Tr}\big[P^K_n (\rho_n - e^{-n\gamma}I_n)\big] \nonumber \\
&= t\, \mathrm{Tr}\big[ P^K_n (\omega_n - e^{-n\gamma}I_n)\big] \nonumber \\
&\phantom{=}\; + (1-t)\, \mathrm{Tr}\big[ P^K_n (\sigma_n - e^{-n\gamma}I_n)\big] \nonumber \\
&\geq t \, \mathrm{Tr}\big[ \{ \omega_n \geq e^{-n\alpha}I_n \} \omega_n \big] \nonumber \\
&\phantom{=}\; + (1-t) \, \mathrm{Tr}\big[ \{ \sigma_n \geq e^{-n\alpha}I_n \} \sigma_n \big] \nonumber \\
&\phantom{=}\;- e^{-n\gamma}\mathrm{Tr}\big[ P^K_n\big] \nonumber \\
&\geq t \, \mathrm{Tr}\big[ \{ \omega_n \geq e^{-n\alpha}I_n \} \omega_n \big] \nonumber \\
&\phantom{=}\; + (1-t) \, \mathrm{Tr}\big[ \{ \sigma_n \geq e^{-n\alpha}I_n \} \sigma_n \big] \nonumber \\
&\phantom{=}\;- 2e^{-n(\gamma - \alpha)} \label{mixed2}
\end{align}
where $\mathrm{Tr}\big[ P^K_n\big] \leq 2e^{n\alpha}$, as the rank of the projector cannot be greater than the sum of the ranks of the projectors $P^0_n$ and $Q_n$.  For every $\delta >
0$ and $\alpha = \max \big[\overline{S}(\sigma), \overline{S}(\omega)\big] +
\delta$, the limit of the sum of first two terms on the RHS goes to 1.  By choosing $\gamma = \alpha + \delta$ this implies both the RHS and LHS converge to 1 and hence that
$$ \overline{S}(\rho) \le \max \big[\overline{S}(\sigma),
\overline{S}(\omega)\big].
$$
as $\delta$ is arbitrary.
\end{proof}

\begin{corollary}
The strong converse is given by
\begin{equation}
\mathcal{R}^* = \min \big[\underline{S}(\sigma), \underline{S}(\omega)\big]
\end{equation}
for any mixed source $\rho_n = t \sigma_n + (1-t) \omega_n$, for $t \in (0,1)$.
\end{corollary}
\begin{proof}
Choosing $\gamma$ and $\alpha$ such that the RHS and LHS of \reff{mixed1} and \reff{mixed2} go to zero, respectively, gives the required inequalities.
\end{proof}

A source obeys the strong converse property only if
\begin{equation}
\underline{S}(\rho) = \lim_{n \rightarrow \infty} \frac{1}{n}S(\rho_n) =
\overline{S}(\rho)
\end{equation}

Note that mixed sources do not obey the strong converse property if $\max \big[
\overline{S}(\sigma), \overline{S}(\omega) \big] > \min \big[
\underline{S}(\sigma), \underline{S}(\omega) \big]$. This can easily be shown to hold
for mixtures of stationary memoryless sources with different entropies
$S(\sigma) > S(\omega)$.

\section{Classical Capacity of an Arbitrary Quantum Channel}
\label{classicalcap}

In this section we obtain the classical capacity of a sequence
of arbitrary quantum channels in terms of the inf-spectral mutual information rate of bipartite separable states shared through the channel.

Let $\{{\cal{K}}_Q^{(n)}\}_{n=1}^\infty$ and
$\{{\cal{H}}_Q^{(n)}\}_{n=1}^\infty$
be two sequences of Hilbert spaces, and
let $\Lambda = \{ \Lambda^Q_n \}_{n=1}^{\infty}$ be a sequence
of quantum channels such that, for each $n$,
$$\Lambda^Q_n : {\cal{B}}({\cal{K}}_Q^{(n)}) \mapsto
{\cal{B}}({\cal{H}}_Q^{(n)}).
$$
Here ${\cal{K}}_Q^{(n)}$ denotes the Hilbert space at the input of the
channel $\Lambda^Q_n$, whereas ${\cal{H}}_Q^{(n)}$ denotes the Hilbert space
at its output.

Consider the following scenario. Suppose Alice has a set of messages,
labelled by the elements of the set ${\cal{M}} = \{1,2, \ldots, M_n\},$
which she would like to communicate to Bob, using the
quantum channel $\Lambda^Q_n$.
To do this, she encodes
each message into a quantum state of a physical system with Hilbert space
${\cal{K}}_Q^{(n)}$ and
sends this state to Bob through the quantum channel.
In order to infer the message that Alice communicated to him, Bob makes
a measurement (described by POVM elements) on the state
that he receives. The encoding and decoding operations together define
a quantum error correcting code (QECC). More precisely, a code
${\cal{C}}^{(n)}$ of size $M_n$ is given by a sequence $\{\rho_n^i,
E_n^i\}_{n=1}^{M_n}$
where each $\rho_n^i$ is a state in ${\cal{B}}({\cal{K}}_Q^{(n)})$ and each
$E_n^i$ is a
positive operator acting in ${\cal{H}}_Q^{(n)}$, such that $\sum_{i=1}^{M_n}
E_n^i \le {\iden}_n$.
Defining $E_n^0 = I_n - \sum_{i=1}^{M_n} E_n^i$, yields a resolution of
identity
in ${\cal{H}}_Q^{(n)}$. Hence, $\{E_n^i\}_{i=0}^{M_n}$ defines a POVM. An
output $i\ge 1$ would
lead to the inference that the state $\rho_n^i$ was transmitted through the
channel $\Lambda^Q_n$,
whereas the output $0$ is interpreted as a failure of any inference. In other
words, a code ${\cal{C}}^{(n)}$ is given by a triple $(M_n, \phi_n, E_n)$,
where
$\phi_n$ is the encoder, i.e., $\phi_n(i) = \rho_n^i$ for $i \in \{1,2,
\ldots, 2^{nR}\}$, and $E_n = \{E_n^i\}_{i=1}^{M_n}$ is the decoder. The
rate of the code is given by $({1}/{n}){\log M_n}$.
The average probability of error for such a code ${\cal{C}}^{(n)}$ is given by
\begin{equation}
P_e({\cal{C}}^{(n)}):= \frac{1}{M_n} \sum_{i=1}^{M_n} \left(1 - {\tr}(\sigma_n^i
E_n^i)\right),
\label{codeerr}
\end{equation}
$\sigma_n^i$ being the output of the channel when the input is the $i^{th}$
codeword $\rho_n^i$. A quantity $R \in {\bf{R}}$ is said to be an
{\em{achievable rate}} if there exists an
$N \in {\bf{N}}$ such that for all $n \ge N$, there exits a sequence
of codes $\{{\cal{C}}^{(n)}\}_{n=1}^\infty$ with $M_n \ge e^{nR}$,
and $P_e({\cal{C}}^{(n)})\rightarrow 0$ as $n \rightarrow \infty$.
\medskip

The capacity of ${\Lambda}$ is defined as
\begin{equation}
C(\Lambda) := \sup R,
\end{equation}
where $R$ is an achievable rate.
%such that there exists a sequence of codes
%$\{{\cal{C}}^{(n)}\}_{i=1}
%\begin{equation}
%\liminf_{n \rightarrow \infty} \frac{1}{n} \log |{\cal{C}}^{(n)}| \ge R
%\label{one}
%\end{equation}
%and
%\begin{equation}
%\lim_{n \rightarrow \infty} P_e({\cal{C}}^{(n)})=0.
%\label{two}
%\end{equation}
%\medskip

%The spectral information rate arising in the
%coding theorem for transmission of classical information
%through a quantum channel is called the {\em{inf-spectral mutual information
%rate}}. For any sequence of bipartite states $\{\rho_n^{XY}\}_{n=1}^\infty$
%it is denoted by ${\underline{S}}(X:Y)$ and defined as follows:
%\be
%\underline{S}(X:Y) = \sup \Big\{ \gamma : \lim_{n\rightarrow \infty}
%\mathrm{Tr}\big[ \{\rho_n^{XY}< \rho_n^X \otimes \rho_n^{Y}\} 
%\rho_n^{XY} \big]= 0 \Big\},
%\ee 
%with $\rho_n^X$,
%$\rho_n^{Y}$ being the reduced density matrices
%of the systems $X$ and $Y$, respectively, corresponding to the state
%$\rho_n^{XY}$.

\begin{theorem}
\label{thm_cap}
The classical capacity of a sequence of channels $\Lambda =
\{ \Lambda^Q_n \}_{n=1}^{\infty}$ is given by
\begin{equation}
C(\Lambda) = \max_{\rho^{AQ} \in \mathcal{S}} \underline{S}(A:\Lambda Q)
\label{statethm}
\end{equation}
where $(i)$ $\mathcal{S}$ is the set of sequences of separable states in
${\cal{B}}({\cal{H}}_{AQ})$, with ${\cal{H}}_{AQ}$ being
a sequence of Hilbert spaces ${\cal{H}}_{AQ} :=\{{\cal{H}}_A^{(n)} \otimes
{\cal{K}}_Q^{(n)}\}_{n=1}^\infty$, and $(ii)$ $\underline{S}(A:\Lambda Q)$ is
the
{\em{inf-spectral mutual information rate}} of a sequence of separable density
matrices $\{\rho_n^{A\Lambda Q}\}_{n=1}^\infty$.
\end{theorem}

Consider an arbitrary set ${\cal{X}}^{(n)}$ of indices and define
a separable state
$$\rho_n^{AQ} := \sum_{x \in {\cal{X}}^{(n)}}p_n^x \rho_{n,x}^A \otimes
\rho_{n,x}^Q,$$
acting in a Hilbert space ${\cal{H}}_A^{(n)} \otimes{\cal{K}}_Q^{(n)}$.
The set of codewords that Alice uses, to transmit her messages to Bob,
is a finite subset of the set
$$\{\rho_{n,x}^Q\,:\, x \in  {\cal{X}}^{(n)}\}.$$
The state $\rho_n^{AQ}$
can be purified to the state
$$\rho_n^{AA'Q} := \sum_{x \in {\cal{X}}^{(n)}}p_n^x |x\rangle \langle x|^{AA'}
\otimes \rho_{n,x}^Q$$
in ${\cal{B}}({\cal{H}}_A^{(n)}\otimes{\cal{H}}_{A'}^{(n)}
\otimes{\cal{K}}_Q^{(n)})$.
Let $B$ denote the bipartite system with Hilbert space
${\cal{H}}_A^{(n)}\otimes{\cal{H}}_{A'}^{(n)}$ (and thus replace
the superscript $AA'$ by $B$). Let $Q$ denote system
with Hilbert space ${\cal{K}}_Q^{(n)}$.
A state $\rho_n^{B Q}$ of the
form given by \reff{rhoq} is referred to as a
{\em{classical--quantum state}}\footnote{
According to the terminology of \cite{windev} it is the density matrix
which one can associate to a {\em{c-q resource}} given by the ensemble
$\{p_n^x, \rho_{n,x}^Q\}$.}. If $X$ is a random variable with
probability mass function $\{p_n^x : x \in {\cal{X}}^{(n)}\}$, then
the state of the quantum system $Q$ is correlated with the values taken by the
classical index $X$. The state $\rho_n^{BQ}$ therefore represents the
preparation of quantum states $\rho_{n,x}^{Q}$ corresponding to classical
indices
$x \in {\cal{X}}^{(n)}$, according to the apriori distribution $\{p_n^x \}$.

The action of the channel $\Lambda^Q_n$ on the system $Q$
yields the state
\bea
\rho_n^{B\Lambda Q} &:=& \bigl({{\hbox{id}}}_{B} \otimes\Lambda^Q_n\bigr)
\bigl(\rho_n^{BQ} \bigr)\nonumber\\
&=& \sum_{x \in {\cal{X}}^{(n)}}p_n^x |x\rangle \langle x|^{B}
\otimes\Lambda^Q_n\bigl( \rho_{n,x}^Q
\bigr)\nonumber\\
&:=& \sum_{x \in {\cal{X}}^{(n)}}p_n^x |x\rangle \langle x|^{B} \otimes
\rho_{n,x}^{\Lambda Q}.
\label{rhoq}
\eea
Here the superscript $\Lambda Q$ is used to denote the system $Q$ after
the action of the channel on it.

For the sequence of classical-quantum states $\{\rho_n^{B\Lambda Q}\}$
the {\em{inf-spectral mutual information rate}} is given by
\begin{equation}
\underline{S}(B:\Lambda Q) = \sup \Big\{ \gamma : \lim_{n\rightarrow \infty}
\mathrm{Tr}\big[ \{ \Pi_n(\gamma) \ge 0\} \Pi_n(\gamma)\big] = 1 \Big\}
\label{def1}
\\
\end{equation}
where $\Pi_n(\gamma) := \rho_n^{B\Lambda Q}- \rho_n^B \otimes \rho_n^{\Lambda
Q}$, and $\rho_n^B$,
$\rho_n^{\Lambda Q}$ are the reduced density matrices
of the systems $B$ and $\Lambda Q$ respectively.

The proof of the Theorem \ref{thm_cap} relies on the following
lemma proved in \cite{hayashi03}.
\begin{lemma}
\label{lemma3HN}
For any $n \in {\bf{N}}$, $M \in{\bf{N}}$,  and
$\gamma \in {\bf{R}}$, given a probability distribution
$\{p^{\ux}_n\}$
on ${\cal{X}}^{(n)}$, there exists
a code ${\cal{C}}^{(n)}$ of size $|{\cal{C}}^{(n)}| = M$, whose average
probability of error satisfies the following bound:
\begin{eqnarray}
P_e({\cal{C}}^{(n)})
&\le & 2\sum_{x \in  {\cal{X}}^{(n)}} p^{x}_n
{{\tr}} \bigl[\{ \rho_{n,x}^{\Lambda Q} - e^{n\gamma} {{\rho}}_n^{\Lambda Q}
\le 0\}
\rho_{n,x}^{\Lambda Q}\bigr]\nonumber\\
& & + 4e^{-n\gamma} M,
\label{lem3}
\end{eqnarray}
where $${{\rho}}^{\Lambda Q}_n:= \sum_{x \in {\cal{X}}^{(n)}} p^{x}_{n}
\rho_{n,x}^{\Lambda Q}.$$
\end{lemma}

\bigskip
{\em{Proof of Theorem \ref{thm_cap}}} We shall first prove that for any rate
$0 < R <\underline{S}(B:\Lambda Q)$, the average probability of error
$P_e({\cal{C}}^{(n)}$ vanishes asymptotically. Here $\underline{S}(B:\Lambda
Q)$
denotes the inf-spectral mutual information rate for a sequence of
{\em{classical-quantum states}} $\{\rho_n^{B \Lambda Q}\}_{n=1}^\infty$
and is given by \reff{def1}.

Computing the reduced density matrices of the bipartite state
$\rho_n^{B\Lambda Q}$
(defined by \reff{rhoq}) yields
\be
\rho^{B}_{n}\otimes \rho^{\Lambda Q}_{n} = \bigl(\sum_{x } p^{x}_{n}
| x\rangle \langle x|^B \bigr) \otimes {{\rho}}_n^{\Lambda Q},
\label{red1}
\ee
where ${{\rho}}_n^{\Lambda Q} := \sum_{x } p^{x}_{n} \rho_{n,x}^{\Lambda
Q}.$
The difference operator $ \Pi_n(\gamma)$ appearing in \reff{def1} is given by
\begin{equation}
\Pi_n(\gamma) = \sum_{x}p^{x}_n |x\rangle \langle x|^B \otimes (\rho^{\Lambda
Q}_{n,x}
- e^{n\gamma}{{\rho}}_n^{\Lambda Q}).
\end{equation}
Note that
\begin{eqnarray}
&&
\mathrm{Tr}\Big[ \{ \Pi_n(\gamma) \ge 0 \}\Pi_n(\gamma)\Big]\nonumber\\
&=&
\mathrm{Tr}\Big[ \{ \Pi_n(\gamma) \ge 0 \} \Bigl(
\sum_{x \in {\cal{X}}^{(n)}} p^{x}_n
| x\rangle \langle x|^B \otimes \bigl(\rho^{\Lambda Q}_{n,x} -
e^{n\gamma}{{\rho}}_n^{\Lambda Q}
\bigr)\Big]\nonumber\\
&& =
\sum_{x} p^{x}_n\mathrm{Tr}
\Bigl[\{\rho^{\Lambda Q}_{n,x} \ge e^{n\gamma} {{\rho}}_n^{\Lambda Q}
\}\bigl(\rho^{\Lambda Q}_{n,x}-e^{n\gamma}{{\rho}}_n^{\Lambda Q}
\bigr)\Bigr].
\end{eqnarray}
Hence, $\underline{S}(B:\Lambda Q)$ is equivalently given by
$$
\sup \Big\{ \gamma : \lim_{n\rightarrow \infty}
\sum_{x} p^{x}_n\mathrm{Tr}
\bigl[\{\rho^{\Lambda Q}_{n,x} \ge e^{n\gamma}{{\rho}}_n^{\Lambda Q}\}
\bigl(\rho^{\Lambda Q}_{n,x} - e^{n\gamma}{{\rho}}_n^{\Lambda Q}
\bigr)\bigr]=1\Big\}.
$$
This implies that for any $\gamma < \underline{S}(B:\Lambda Q)$,
\begin{equation}\lim_{n\rightarrow \infty}\sum_{x} p^{x}_n\mathrm{Tr}
\bigl[\{\rho^{\Lambda Q}_{n,x} < e^{n\gamma}{{\rho}}_n^{\Lambda Q}\}
\rho^{\Lambda Q}_{n,x}\bigr] = 0.
\label{eq2}
\end{equation}
For $M_n = \lceil e^{nR} \rceil$, Lemma \ref{lemma3HN}
ensures the existence of a sequence of codes
$\{{\cal{C}}^{(n)}\}_{n=1}^\infty$ of size $|{\cal{C}}^{(n)}| = \lceil e^{nR}
\rceil$, such that for each $n$
\begin{eqnarray}
P_e({\cal{C}}^{(n)})
&\le & 2\sum_{x \in  {\cal{X}}^{(n)}} p^{x}_n
{{\tr}} \bigl[\{ \rho_{n,x}^{\Lambda Q} - e^{n\gamma} {{\rho}}_n^{\Lambda Q}
\le 0\}
\rho_{n,x}^{\Lambda Q}\bigr]\nonumber\\
& & + 4e^{-n\gamma} \lceil e^{nR} \rceil,
\label{eq22}
\end{eqnarray}
for any $\gamma \in {\bf{R}}$ and $c >0$.
From (\ref{eq2}) it follows that for any $\gamma < \underline{S}(B:\Lambda Q)$,
the first term on the RHS of (\ref{eq22}) vanishes in the
limit $n \rightarrow \infty$. 
For all $\delta >0$, there exists $n_0 \in {\bf{N}}$, such that for all $n \ge n_0$, $\lceil e^{nR} \rceil \le e^{n(R + \delta)}$. Hence,
$$ 4e^{-n\gamma}  \lceil e^{nR} \rceil
\leq 4^{-n(\gamma- (R + \delta))},$$
which vanishes as $n \rightarrow \infty$ for $\gamma > R + \delta$. Since 
$\delta$ is arbitrary, it follows that any rate $R < \gamma 
<\underline{S}(B:\Lambda Q)$ is achievable. More generally, any rate
$0 < R <\underline{S}(B:\Lambda Q)$ is achievable.

% Now we proceed to prove the (weak) converse, i.e., for any rate $R >
% {\underline{S}}(B:\Lambda Q)$ there does not exist any sequence of codes
% $\{{\cal{C}}^{(n)}\}_{n=1}^\infty$ whose average probability of error vanishes
% asymptotically. For any code
% ${\cal{C}}^{(n)}$ of rate $R:= \frac{1}{n} \log {M_n}$,
% Lemma \ref{lemma4HN} implies that
% \begin{eqnarray}
% P_e({\cal{C}}^{(n)})
% &\geq & \sum_{x \in {\cal{X}}^{(n)}} p^{x}_n
% {{\tr}} \bigl[\{ \rho_{n,x}^{\Lambda Q} < e^{n\gamma} {{\rho}}_n^{\Lambda
% Q}\}
% \rho^{\Lambda Q}_{n,x}\bigr]\nonumber\\
% & & - e^{-n(R-\gamma)}.
% \label{lem4a}
% \end{eqnarray}

To prove the (weak) converse we are required only to show that for any code with rate larger than the capacity, there exists a probability distribution on the codewords such that the average probability of error does not vanish asymptotically.

Define a family of codes of size $M_n$ by the average state of the codewords $\rho^Q_n$.  Note that the family includes all possible sets of $M_n$ codewords with the same average state.  Given the family $\{ M_n, \rho^Q_n \}_{n=1}^{\infty}$ we can extend $\rho^Q_n$ to any separable state $\rho^{AQ}$ on an enlarged Hilbert space.  The outcome of any measurement on $A$ is thus classically correlated with a state on $Q$.

Explicitly, we can assign the message that has been sent with the outcome of the set of measurements on $A$, described by a POVM $\{E_{n,i}^A\}$, such that message $i \in \{ 1,2,..., M_n \}$ is generated with probability
\begin{equation}
p_i = \mathrm{Tr}\big[ (E_{n,i}^A \otimes I^Q_n) \rho^{AQ}_n \big].
\label{prob}
\end{equation}
and results in the codeword
\begin{equation}
\rho^Q_{n,i} = \mathrm{Tr}_A \big[ \sqrt{E_{n,i}^A \otimes I^Q_n} \rho^{AQ}_n \sqrt{E_{n,i}^A \otimes I^Q_n}\big]
\end{equation}
which is then sent throught the noisy channel.

\noindent
The average probability of error can thus be expressed as
\begin{align}
P_e({\cal{C}}^{(n)}) &= 1 - \sum_{i=1}^{M_n} p_i {\tr}\big[ E^{Q}_{n,i}\Lambda^Q_n \rho^Q_{n,i})\big] \nonumber \\
&= 1 - \sum_{i=1}^{M_n} {\tr}\big[ (E^A_{n,i}\otimes E^{Q}_{n,i}) \rho^{A\Lambda Q}_n\big],
\end{align}
where $\rho^{A\Lambda Q}_n = (I^A_n \otimes \Lambda^Q_n)\rho^{AQ}_n$.

From Lemma \ref{lemma} it then follows that
\begin{align}
P_e({\cal{C}}^{(n)}) &\geq 1 - \mathrm{Tr}\big[ \{ \Pi_n(\gamma) \geq 0 \} \Pi_n (\gamma) \big] \nonumber \\
&\phantom{=}\;- e^{n\gamma}\mathrm{Tr}\big[ \sum_{i=1}^{M_n} E^A_{n,i}\rho^{A}_n \otimes E^{Q}_{n,i} \rho^{\Lambda Q}_n\big] \nonumber \\
&= 1 - \mathrm{Tr}\big[ \{ \Pi_n(\gamma) \geq 0 \} \Pi_n (\gamma) \big] \nonumber \\
&\phantom{=}\;- e^{n\gamma}\sum_{i=1}^{M_n} p_i \mathrm{Tr}\big[ E^{Q}_{n,i} \rho^{\Lambda Q}_n\big]
\end{align}
with $\Pi_n (\gamma) = \rho^{A\Lambda Q}_n - e^{n\gamma}\rho^{A}_n \otimes \rho^{\Lambda Q}_n$, and where the probability $p_i$ is given by \reff{prob}.

Choosing only those POVMs such that
\begin{equation}
p_i = \mathrm{Tr}\big[ E_{n,i}^A  \rho^{A}_n \big] = \frac{1}{M_n}
\end{equation}
is sufficient to show that any code of size $M_n$ is not reliable.  In this case
\begin{equation}
P_e({\cal{C}}^{(n)}) \geq 1 - \mathrm{Tr}\big[ \{ \Pi_n(\gamma) \geq 0 \} \Pi_n (\gamma) \big] - \frac{e^{n\gamma}}{M_n}
\label{lem4a}
\end{equation}
and for any $\delta >0$, choose $M_n = \lceil e^{nR} \rceil$ where $R = \underline{S}(A:\Lambda Q) + 2\delta$,
and $\gamma = \underline{S}(A:\Lambda Q) + \delta$. Thus, the third term on the
RHS of \reff{lem4a} vanishes in the limit $n \rightarrow \infty$. However,
the difference of the first two terms does not vanish and we have
$\limsup_{n\rightarrow \infty} P_e({\cal{C}}^{(n)}) \ge \epsilon_0$ for some $\epsilon_0 >0$.

We thus conclude that the classical capacity of a
sequence of channels $\Lambda =
\{ \Lambda^Q_n \}_{n=1}^{\infty}$ is given by
\begin{equation}
C(\Lambda) = \max_{\rho^{BQ} \in \mathcal{Q}} \underline{S}(B:\Lambda Q)
\end{equation}
where $\mathcal{Q}$ denotes the set of sequences of classical--quantum
states in ${\cal{B}}({\cal{H}}_{BQ})$, with ${\cal{H}}_{BQ}$ being
a sequence of Hilbert spaces ${\cal{H}}_{BQ} :=\{{\cal{H}}_B^{(n)} \otimes
{\cal{K}}_Q^{(n)}\}_{n=1}^\infty$,
The monotonicity of
the inf-spectral mutual information rate under CPTP maps
(see \cite{BD1}) implies that
$\underline{S}(B:\Lambda Q) \equiv \underline{S}(AA':\Lambda Q)
\geq \underline{S}(A:\Lambda Q)$.
This ensures that optimization over classical-quantum states is equivalent
to optimization over separable states, thus yielding the statement
\ref{statethm} of Theorem \ref{thm_cap}.

% \subsection{Mixed Channels}
% 
% A mixture of channel is any finite convex sum of channels, such that
% \begin{equation}
% \tilde{\Lambda} = \sum_i q_i \Lambda^{(i)}
% \end{equation}
% where $\sum_i q_i = 1$.

%%% Insert Dense Coding %%%%%%%%%%%%
\section{Dense Coding}

Dense coding is the protocol by which prior shared entanglement between
a sender (Alice) and a receiver (BOB) is exploited for sending classical
messages through a noiseless quantum channel. Let ${\rho_n^{AB}}\in
{\cal{H}}_A^{(n)} \otimes {\cal{H}}_B^{(n)} $ be an entangled mixed
state that Alice and Bob initially share. As in Section \ref{classicalcap},
Alice has a set of messages,
labelled by the elements of the set ${\cal{M}}_n = \{1,2, \ldots, M_n\},$
which she wishes to communicate to Bob. However, the quantum channel
that she uses is noiseless. She encodes her messages into her part, $A$, of
the bipartite system $AB$ which is in the state ${\rho_n^{AB}}$. The codewords
are given by
$$\phi_n(i):=\rho_{n,i}^{AB} = ({\cal{E}}_{n,i}^A \otimes
{\hbox{id}}^B)\rho_{n}^{AB},$$
for $i= {\cal{M}}_n$. Here $\phi_{n}$ denotes the encoding map
for a code of size $M_n$ as defined in terms of the CPTP maps
${{\cal{E}}}_{n,i}^A$, $i \in {\cal{M}}_n$.
Let Bob's measurement on the states $\rho_{n,i}^{AB}$
that he receives, be given by
$E_n^{AB} =\{E_{n,i}^{AB}\}_{i=1}^{M_n}$, with each $ E_{n,i}^{AB}\ge 0$ and
$\sum_{i=1}^{M_n} E_{n,i}^{AB} \le I^{AB}_n$. The average probability of
error of the code ${\cal{C}}^{(n)}=(M_n, \phi_n^A, E_n^{AB})$ is given by
\begin{equation}
P_e({\cal{C}}^{(n)}):= \frac{1}{M_n} \sum_{i=1}^{M_n} \left(1 -
{\tr}(\rho_{n,i}^{AB}
E_{n,i}^{AB})\right),
\label{codeerr2}
\end{equation}
The dense coding capacity for a sequence of bipartite states
$\rho^{AB} = \{ \rho^{AB}_n \}_{n=1}^{\infty}$ is defined as
\begin{equation}
C_{DC} := \sup R,
\end{equation}
where $R$ is an achievable rate.
\medskip

\begin{theorem}
The dense coding capacity for a sequence of bipartite states $\rho^{AB} = \{
\rho^{AB}_n \}_{n=1}^{\infty}$ is given by
\begin{equation}
C_{DC} = \log d - \min_{\Lambda} \overline{S}(\Lambda A |B)
\label{cdc}
\end{equation}
where $\Lambda = \{ \Lambda_n^A \}_{n=1}^{\infty}$ is a sequence of CPTP maps
on $A$.
\end{theorem}

\begin{proof}[Converse]
For a code ${\cal{C}}^{(n)}$ of $M_n$ codewords
$\rho_{n,i}^{AB} = (\phi_{n,i}^A \otimes {\hbox{id}}^B)\rho_{n}^{AB}$, and
measurement operators $E_{n,i}^{AB}$, $i=1, \ldots, M_n$, the average
probability of error \reff{codeerr2} satisfies
\begin{align}
P_e({\cal{C}}^{(n)})
&\geq 1 - \frac{1}{M_n} \sum_i \mathrm{Tr}\big[E_{n,i}^{AB}
\rho_{n,i}^{AB} - e^{-n\gamma} I_n^A \otimes \rho^{B}_n\big]\nonumber\\
&-
\frac{e^{-n\gamma}}{M_n}\mathrm{Tr}\Big[ E_{n,i}^{AB}
(I_n^A \otimes \rho^{B}_n) \Big] \nonumber \\
&\geq 1 - \frac{1}{M_n} \sum_i \mathrm{Tr}\big[ \Pi_{n,i}(\gamma) \big] -
\frac{e^{-n\gamma}}{M_n}\mathrm{Tr}\,I_n^A \nonumber \\
&\geq
1 - \max_i \mathrm{Tr}\big[ \Pi_{n,i}(\gamma) \big] - \frac{e^{n(\log d -
\gamma)}}{M_n}
\end{align}
where $\Pi^i_n(\gamma) = \{  \rho_{n,i}^{AB}\geq e^{-n\gamma} I_n^A
\otimes \rho^{B}_n \} \big( \rho_{n,i}^{AB} - e^{-n\gamma} I_n^A \otimes
\rho^{B}_n \big)$. In the above we have used Lemma \ref{lemma} and the facts
that $\sum_i E_{n,i}^{AB} \le I_n^{AB}$ and ${{\tr}\,{I^A_n}}= e^{n \log d}$.

If we then assume that $M_n \geq e^{nR} = \log d - \min_{\Lambda}
\overline{S}(\Lambda A|B) + 2\delta$ for some $\delta > 0$,
then we can choose $\gamma = \min_{\phi} \overline{S}(\phi A|B) - \delta$,
and we find
\begin{equation}
\limsup_{n\rightarrow \infty} P_e({\cal{C}}^{(n)})
\geq \epsilon_0 > 0
\end{equation}
implying $C_{DC} \leq \log d - \min_{\phi} \overline{S}(\phi A|B)$.
\end{proof}

\begin{proof}[Coding]
%%%INSERT %%%
Lemma \ref{lemma3HN}, adapted to the case of dense coding, states that
for any $n \in {\bf{N}}$, $M \in{\bf{N}}$,  and
$\gamma \in {\bf{R}}$, given a probability distribution
$\{p^{\ux}_n\}$
on ${\cal{X}}^{(n)}$, where ${\cal{X}}^{(n)}$ is a finite set of
indices, there exists
a code ${\cal{C}}^{(n)}$ of size $|{\cal{C}}^{(n)}| = M$, whose average
probability of error satisfies the bound
\begin{eqnarray}
P_e({\cal{C}}^{(n)})
&\le & 2\sum_{x \in  {\cal{X}}^{(n)}} p^{x}_n
{{\tr}} \bigl[\{ \rho_{n,x}^{AB} < e^{n\gamma} {{\rho}}_n^{AB} \}
\rho_{n,x}^{AB}\bigr]\nonumber\\
& & + 4e^{-n\gamma} M,
\label{lem33}
\end{eqnarray}
where $${{\rho}}_n^{AB}:= \sum_{x \in  {\cal{X}}^{(n)}} p^x_n
\rho_{n,x}
^{AB}.$$
%{\bf{The following bit needs to be improved}}
Choose ${\cal{X}}^{(n)}$ to be a set of size $N_n= d^{2n}$
and define a probability distribution $\{p_n^x\}$ on it,
with $p_n^x = 1/{N_n} = e^{-2n \log d}$ for each $x \in  {\cal{X}}^{(n)}$.
Further, consider states $\rho_{n,x}^{AB}$ defined as follows:
$$\rho_{n,x}^{AB} := \bigl(\mathcal{U}_{n,x}^A {{\Lambda}}_{n}^A
\otimes {\hbox{id}}^B)\bigr)\rho_{n}^{AB}.$$
Here ${{\Lambda}}_{n}^A$ denote quantum operations for which the
sequence $\{{{\Lambda}}_{n}^A\}_{n=1}^\infty$ minimizes
$\overline{S}(\Lambda A|B)$, and $(ii)$ $\mathcal{U}_{n,x}^A,$ $x \in
{\cal{X}}^{(n)}$, denotes
unitary encodings with the shift-multiply operators
$U_{(p,q)}$, with $p,q \in \{0,1,\ldots ,(d^n -1) \}$,
which are defined as follows (\cite{hiroshima, bowen1}):
$$U_{(p,q)}|j\rangle = e^{\frac{2\pi p j}{d}}|j + q\, ( \textrm{mod}\, d)\rangle,$$
with $\{|j\rangle :j \in \{0,1,\ldots ,(d^n -1) \}$ being an orthonormal
basis in a $d^n$-dimensional Hilbert space.

Let $$\rho_{n}^{\Lambda AB} :=
({\Lambda}_{n}^A \otimes {\hbox{id}}^B)\rho_{n}^{AB}.$$
For the ensemble $\{p^x_n, \rho_{x,n}^{AB}\}$
\begin{eqnarray}
\sum_{x \in {\cal{X}}^{(n)}} p^x_n
\rho_{n,x}^{AB}
&=&\sum_{x \in {\cal{X}}^{(n)}} p^x_n \bigl(\mathcal{U}_{n,x}^A
\otimes {\hbox{id}}^B)\rho_{n}^{\Lambda AB}\nonumber\\
&=& \frac{I^A_n}{d^n} \otimes \rho^B_n,
\end{eqnarray}
where $\rho^B_n$ is the reduced density matrix of the state
$\rho_n^{\Lambda AB}$.
\medskip

For the ensemble $\{p^x_n, \rho_{x,n}^{AB}\}$ defined above, let
$$\alpha_n:= \sum_{x \in  {\cal{X}}^{(n)}} p^{x}_n
{{\tr}} \bigl[\{ \rho_{n,x}^{AB} \ge e^{n\gamma} {{\rho}}_n^{AB} \}
\rho_{n,x}^{AB}\bigr]$$
We have that
\begin{align}
\alpha_n &\geq \frac{1}{N_n}\sum_{x \in  {\cal{X}}^{(n)}}
{{\tr}} \bigl[\{ \rho_{n,x}^{AB} \ge e^{n\gamma} {{\rho}}_n^{AB} \} \nonumber\\
&\phantom{=}\; \times \bigl(\rho_{n,x}^{AB} - e^{n\gamma} {{\rho}}_n^{AB} \bigr)
\bigr]\nonumber\\
&= {{\tr}} \bigl[\{ \rho_{n}^{\Lambda AB} \ge e^{-n(\log d -\gamma)}
I^A_n \otimes \rho^B_n \} \nonumber\\
&\phantom{=}\;\times \bigl(\rho_{n}^{\Lambda AB} - e^{-n(\log d - \gamma)} {I^A_n}\otimes \rho^B_n \bigr)
\bigr].\label{dc}
\end{align}
In the above we have made use of the fact that the trace remains invariant
under a unitary transformation.
If $\gamma = \log d -
\overline{S}({\Lambda} A|B) -
\delta$ for any $\delta > 0$, the RHS of \reff{dc} goes to one as
$n \rightarrow \infty$. Hence the RHS of \reff{lem33} vanishes asymptotically,
implying that a rate
$R = \log d - \min_{\Lambda} \overline{S}(\Lambda A|B) - \delta$ is achievable for any $\delta > 0$.
\end{proof}

\subsection{Reduction to the {i.i.d.} Case}
For entanglement resources which are tensor products of identical bipartite 
states $\rho^{AB}_N = \varrho_{AB}^{\otimes N}$, with $\varrho_{AB}
\in {\cal{B}}({\cal{H}})$, the dense coding capacity was shown in \cite{horodecki} to be given by 
\begin{equation}
C_{DC} = \log d + S(B) - \inf_N \inf_{\Lambda_A^{(N)}} \frac{1}{N} 
S\big( (\Lambda_A^{(N)}\otimes {\hbox{id}}_B^{(N)})\varrho_{AB}^{\otimes N} \big).
\label{hor}
\end{equation}
Here $S(B)=S(\varrho_B)$, where $\varrho_B$ is the reduced density
matrix of the system $B$, corresponding to the state $\varrho_{AB}$.

For sequences of {i.i.d.} states 
${{\omega}} = \{ \vartheta^{\otimes n} \}_{n=1}^{\infty}$ and 
${\sigma} = \{ \varsigma^{\otimes n} \}_{n=1}^{\infty}$, Theorem 4 of 
\cite{nagaoka02} 
states that
\begin{equation}
\underline{D}(\omega\| \sigma) = D(\vartheta \| \varsigma ) = \overline{D}(\omega \| \sigma).
\label{thm4}
\end{equation}
For bipartite states $\vartheta= \vartheta_{AB}$ and $\varsigma= 
\tfrac{1}{d}I_A \otimes \vartheta_B$, \reff{thm4} implies that 
\begin{equation}
\overline{S}(A|B) = S(A|B) = \underline{S}(A|B),
\end{equation}
where $S(A|B) = S(\vartheta^{AB}) - S(\vartheta^B)$.
This is because $\log d - \overline{S}(A|B) = \underline{D}(\omega \| \sigma) 
= D(\vartheta \| \varsigma ) = \log d - \frac{1}{n}
S\bigl(\vartheta_{AB}^{\otimes n}|\vartheta_{B}^{\otimes n}\bigr) = \log d - S(A|B)$, and similarly for 
$\overline{D}(\omega \| \sigma)$. If instead, we choose 
$\vartheta_{AB}$ and $\varsigma_{AB}$ to be states in ${\cal{B}}({\cal{H}}^{\otimes N})$, given by
$$
\vartheta_{AB} := \bigl(\Lambda_A^{(N)}\otimes {\hbox{id}}_B^{(N)} \bigr)
\varrho_{AB}^{\otimes N},$$ 
and 
$$\varsigma_{AB} = \tfrac{1}{d^N}I_A^{(N)} \otimes \varrho_B^{\otimes N},$$
then \reff{thm4} yields the identity
\begin{equation}
\overline{S}(\Lambda^{(N)} A|B) = \frac{1}{N}S\Bigl((\Lambda_A^{(N)}\otimes 
{\hbox{id}}_B^{N})\varrho_{AB}^{\otimes N}\Bigr) - S(\varrho_B).
\end{equation} 
Hence, in this case our expression \reff{cdc} for the dense coding capacity 
reduces to \reff{hor}.
%\end{theorem}

\end{document}